\documentclass[11pt,twoside]{article}

\usepackage{asp2006}
\usepackage{epsf}
\usepackage{psfig}
\usepackage{lscape}
\usepackage{graphicx}
\usepackage{subfigure}

\begin{document}
\title{A New Lower Limit to the Galactic Centre's Magnetic Field}   
\author{Roland M. Crocker}   
\affil{Max-Planck-Institut f{\" u}r Kernphsik,\\ 
P.O. Box 103980 Heidelberg, 
Germany}    

\begin{abstract} 
The amplitude of the magnetic field surrounding the Galactic Centre (GC) on large scales ($> 100$pc) has been
uncertain by two orders of magnitude for several decades:
different analyses report fields as weak as $\sim$6 $\mu$G on the one hand and $\sim$1 mG on the other.
Here I report on our recent work \citep{Crocker2010} which shows that the field on  400 pc scales has a firm lower limit of about 50 $\mu$G.
To obtain this result we compiled existing (mostly single dish) radio data to construct the spectrum of the GC region on these size scales.
This spectrum is a broken power law with a down-break (most conservatively)
attributable to a transition from bremsstrahlung to
synchrotron cooling of the in-situ cosmic-ray electron population.
The lower limit on the magnetic field arises through the consideration that the synchrotron-emitting electrons should not
produce too much $\gamma$-ray emission given existing constraints from the EGRET instrument. 
\end{abstract}



There has long been controversy regarding the strength of the Galactic centre's ambient, large-scale magnetic field.
On a scale of
$\sim$100 pc, fields of $\sim$1000 $\mu$G have been inferred from observations of the
mysterious non-thermal radio filaments \citep{Yusef-Zadeh1987,Morris1989,Morris2007} .
Alternatively, the assumption of
pressure equilibrium between the various phases of the Galactic Centre
interstellar medium (including turbulent molecular gas; the
contested\citep{Revnivtsev2009} ``very hot" plasma;
and the magnetic field) suggests fields of $\sim$100 $\mu$G\citep{Spergel1992}.
Finally, radio observations at 74 MHz and 330 MHz\citep{LaRosa2005} reveal a diffuse region of non-thermal radio emission (`{\bf DNS}') 
covering the GC (out to $\sim \pm3^\circ$ or $\sim \pm420$ pc from the GC along the Galactic plane).  
Invoking
the equipartition condition,  a field of only 6 $\mu$G is inferred\citep{LaRosa2005} from these observations, 
typical for the Galaxy-at-large (for extremal parameter values  the field might reach 100 $\mu$G). 

To probe this  radio structure at higher frequencies, 
we assembled archival total-intensity, single-dish flux density data at 1.4, 2.4, 2.7, and 10 GHz \newline \citep{Reich1990, Reich1984,Duncan1995,Handa1987}.  
These data were processed to remove the pollution due to line-of-sight synchrotron emission 
in the Galactic plane both behind and in front of the GC (measured in 408 MHz all-sky Galactic synchrotron background observations \citep{Haslam1982}) and  to remove the flux density attributable to discrete sources in the field (via spatial wavenumber filtering).
   
After this processing a distinct, non-thermal, radio structure is revealed in all radio maps up to 10 GHz. 
However, the structure's 74 MHz--10 GHz 
spectrum is not described by a pure power-law: 
this provides a poor fit with a minimum  $\chi^2$ of 4.9 per degree of freedom ($dof = 4$), excluded at a confidence level of $3.4\sigma$. 

\begin{figure}
\renewcommand \thefigure{1}
\includegraphics{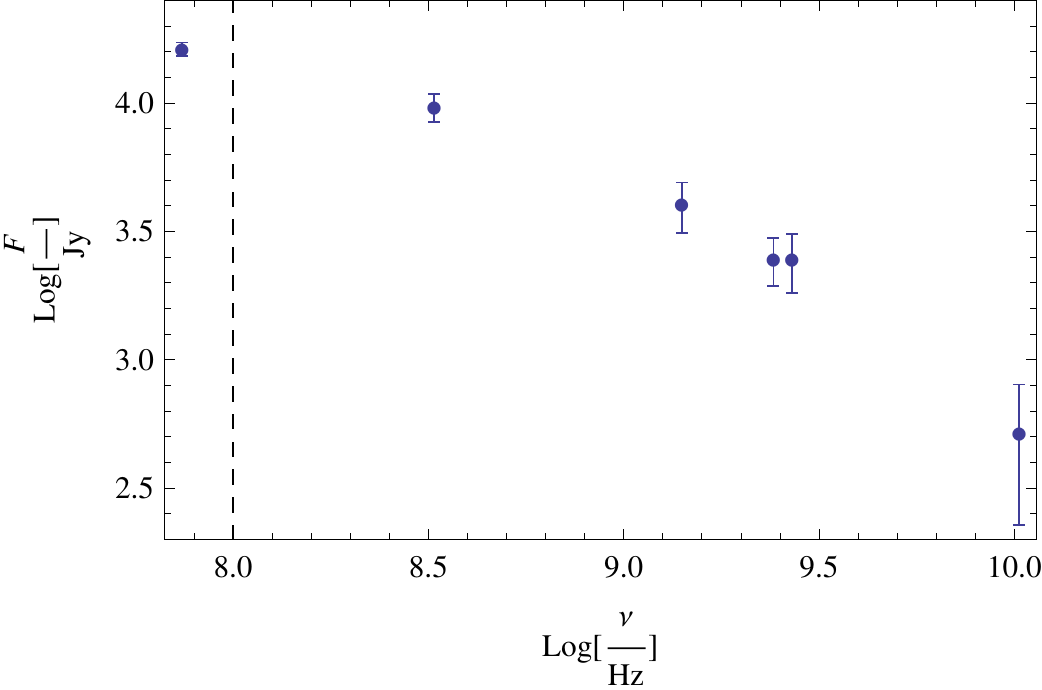}
\caption{DNS radio spectrum. The DNS is described by an ellipse, centred on the GC, of $3^\circ$ semi-major axis and $1^\circ$ semi-minor axis\citep{LaRosa2005}.  The vertical dashed line divides the 74 MHz datum (which receives no contribution from the Galactic plane background/foreground) from the other data (which do). The break is just above $\sim$GHz.}
\label{fig_spctrm}
\end{figure}

On the other hand, fitting separate power laws to, respectively, the background-
subtracted lower three and upper three radio data, we find that these extrapolations intersect at $\sim$ 1.7 GHz with a spectral down-break of $\sim$0.6. This is close to the canonical break of 1/2 produced by a steady-state
synchrotron-radiating electron population that transitions (with increasing energy) from 
bremsstrahlung-cooled to synchrotron-cooled.  
 
Given these data, we may place a lower limit on the ambient magnetic field as we now explain. 
The logic of this is the following: for any given $B$, given the detection of the break, one then knows $n_H$ (since the synchrotron cooling rate is a function of magnetic field, $B$, and bremsstrahlung a function of $n_H$,  the break frequency is a function of both parameterss). Moreover, a certain synchrotron luminosity is observed, so for any given $B$, the ambient cosmic ray electron population is also known. Thus, for any given $B$, one knows the $n_H$ value and the relativistic electron population (for the region where the synchrotron emission is occurring) and one can therefore predict the amount of bremsstrahlung emission from this population. (Inverse Compton emission and cooling processes with ambient light\citep{Porter2006} are also taken into account in our modelling, as is ionization cooling). As one `dials down' the ambient $B$ field, so many ambient cosmic rays electrons are eventually required (to maintain the observed synchrotron luminosity) that they produce too much bremsstrahlung (and IC) emission to obey the 300 MeV $\gamma$-ray flux upper limit from the region measured by EGRET\citep{Hunter1997}. 
This consideration rules out field amplitudes $<$ 50 $\mu$G. (We eagerly await results from the Fermi Gamma-ray Space Telescope\citep{Atwood2009} on the GC region.)
 
From our modelling we find that  the cosmic ray electron population must be considerably sub-equipartition with respect to the other GC ISM phases
explaining why the magnetic field estimate arrived at 
assuming equipartition \citep{LaRosa2005} is too low. 
In contrast, at $\sim$ 100 $\mu$G the magnetic field reaches equipartition (at $\sim$300 eV cm$^{-3}$) with the putative (and contested \citet{Revnivtsev2009}) `very hot' ($\sim$ 8 keV) phase of the X-ray-emitting plasma supposedly detected throughout the central few degrees along the Galactic plane \citep{Koyama1989}.
The energy density of the gas turbulence kinetic energy for the derived $n_H$ is also within a factor of a few of equipartition with these other phases for magnetic field amplitudes up to $\sim 100 \  \mu$G.
Thus, 
as originally speculated by Spergel and Blitz \citep{Spergel1992}, the GC ISM thus seems to be characterised by near
pressure equilibrium between a number of its phases (including the 8 keV plasma -- if it is real), mirroring the situation for the Galactic disk (albeit at much higher pressure). 

Also noteworthy is a number of interesting parallels between the GC situation and that apparently pertaining within the inner regions of starburst galaxies. 
In starburst environments, 
Thompson et al.\citep{Thompson2006} contend that  equipartition magnetic field values significantly underestimate real field amplitudes which are, instead, supposed to be high enough to be in (or close to) hydrostatic equilibrium with the self-gravity of the gaseous disk of the starburst. 
Finally, Thompson et al.\citep{Thompson2006} also speculate that  radio emission from starbursts is dominated by secondary electrons created in collisions between cosmic ray ions and gas (rather than directly-accelerated electrons).

It is noteworthy, then, that 
a $\sim$100 $\mu$G field is precisely in the range required to establish hydrostatic equilibrium in the GC (given the total and gaseous surface densities inferred from the data presented in \citep{Ferriere2007}). 
Moreover, a situation wherein a magnetic field provides significant pressure support against gravity may lead to the development of the Parker instability \citep{Parker1966}
and exactly this is suggested by mm-wave observations \newline \citep{Fukui2006} of molecular filaments of several hundred parsec length within $\sim$ 1 kpc of the GC. These observations independently suggest a $\sim$ 100 $\mu$G field.

Finally, the diffuse, $\sim$TeV $\gamma$-ray glow from the vicinity of the GC \newline  \citep{Aharonian2006} is most likely explained by cosmic ray impacts with gas. Unavoidably, such collisions would also produce copious secondary electrons which could then contribute significantly to the region's synchrotron radio emission (for fields $> 300 \mu$G, 100\% of the radio emission from the region defined by $|l| < 0.8^\circ$ and $|b| < 0.3^\circ$ -- the region over which HESS detects diffuse $\sim$TeV emission -- could be attributed to secondary electrons).
Taken altogether, these facts paint the Galactic centre as akin to a weak starburst with a magnetic field of $\sim$100 $\mu$G.

\acknowledgements 

I gratefully acknowledge the input of my co-authors \citep{Crocker2010} and collaborators: David Jones, Fulvio Melia, \ J{\" u}rgen Ott, \& Raymond J. Protheroe.



\end{document}